\documentclass[superscriptaddress,preprint,amsmath,amssymb,aps,prl]{revtex4-2}

\usepackage{graphicx}
\usepackage{amsmath}
\usepackage{dcolumn}
\usepackage{txfonts}
\usepackage{bm}
\usepackage{mathrsfs}
\usepackage{amsfonts}
\usepackage{amssymb}
\usepackage{braket}
\usepackage{color}
\usepackage[colorlinks=true,citecolor=magenta,anchorcolor=blue]{hyperref}

\begin{document}

\title{Non-Hermitian physics and topological phenomena in convective thermal metamaterials}

\author{Zhoufei Liu}\email{zfliu20@fudan.edu.cn}
\affiliation{Department of Physics, State Key Laboratory of Surface Physics, and Key Laboratory of Micro and Nano Photonic Structures (MOE), Fudan University, Shanghai 200438, China}

\date{\today}

\begin{abstract}

Non-Hermitian physics and topological phenomena are two hot topics attracted much attention in condensed matter physics and artificial metamaterials. Thermal metamaterials are one type of metamaterials that can manipulate heat on one's own. Recently, it has been found that non-Hermitian physics and topological phenomena can be implemented in purely diffusive systems. However, conduction alone is not omnipotent due to the missing of degrees of freedom. Heat convection, accompanying with conduction, is capable of realizing a large number of phases. In this review, we will present some important works on non-Hermitian and topological convective thermal metamaterials. In non-Hermitian physics, we will first discuss the implementation of exceptional point (EP) in thermal diffusion, followed by high-order EP and dynamic encirclement of EP. We then discuss two works on the extensions of EP in diffusion systems, namely, the chiral thermal behavior in the vicinity of EP and the Weyl exceptional ring. For topological phases, we will discuss two examples: a one-dimensional topological insulator and a two-dimensional quadrupole topological insulator. Finally, we will make a conclusion and present a promising outlook in this area. Besides the scientific values, non-Hermitian and topological convective thermal metamaterials have great potentials for industrial applications.   

\end{abstract}

\maketitle

\section{${\rm{\textbf{\uppercase\expandafter{\romannumeral1}}}}$. Introduction}

The effective non-Hermitian Hamiltonian provides a simple and intuitive method to understand the dynamics of open systems~\cite{ZFLiu-AshidaAP20, ZFLiu-BergholtzRMP21}. In recent decades, the study of non-Hermitian physics has been a frontier and attracted many researchers from condensed matter physics, optics, quantum information, etc. Among this field, parity-time (PT) symmetry is one of the most important symmetries because the system will exhibit a real spectrum~\cite{ZFLiu-BenderRPP07, ZFLiu-BenderPRL98}. When the eigenstate of PT-symmetric Hamiltonian preserves (violates) PT symmetry, the system is in the PT unbroken (broken) phase. The critical point between these two phases is called the exceptional point (EP). Besides, many interesting phenomena without the Hermitian counterpart are proposed in non-Hermitian physics, such as non-Hermitian skin effect~\cite{ZFLiu-YaoPRL18, ZFLiu-OkumaPRL20, ZFLiu-ZhangPRL20} and non-Hermitian topological classification~\cite{ZFLiu-GongPRX18, ZFLiu-KawabataPRX19}.

The quantum Hall effect discovered in the 1980s is the first proposed topological phase transition beyond the conventional Landau paradigm~\cite{ZFLiu-KlitzingPRL80, ZFLiu-Klitzing05}. In the last twenty years, it witnessed the flourishing of topological physics, and plenty of topological phases of matter have been theoretically predicted and experimentally discovered~\cite{ZFLiu-HasanRMP10, ZFLiu-QiRMP11}. Apart from condensed matter physics, topology has been incorporated into classical wave systems, giving rise to many new fields. One most typical and significant example is topological photonics~\cite{ZFLiu-LuNP14, ZFLiu-OzawaRMP19}. The basic principle for topological photonics is the mapping between Maxwell's equations in electrodynamics and Schr${\rm{\ddot{o}}}$dinger equation in quantum mechanics. Other examples, such as topological acoustics~\cite{ZFLiu-YangPRL15, ZFLiu-XueNRM22} and topological mechanics~\cite{ZFLiu-HuberNP16}, have also been good platforms for realizing topological states.  

Thermal metamaterials have become the focus of growing interest in physics, materials science, and thermophysics~\cite{ZFLiu-Huang20, ZFLiu-Xu23, ZFLiu-YangPR21, ZFLiu-ZhangNRP23, ZFLiu-YangRMP23, ZFLiu-WangiScience20, ZFLiu-HuangPhysics20, ZFLiu-HuangESEE20, ZFLiu-HuangESEE19, ZFLiu-HuangPP18, ZFLiu-JiIJMPB18, ZFLiu-Tan20, ZFLiu-LiNRM21, ZFLiu-JuAM22}. Based on transformation thermotics, thermal cloaking was first proposed in 2008~\cite{ZFLiu-FanAPL08}. After that, more thermal functions with different artificial structures became possible, such as thermal illusion~\cite{ZFLiu-XuEPJB19-1, ZFLiu-YangESEE19, ZFLiu-WangIJTS18, ZFLiu-XuJAP18, ZFLiu-XuPLA18, ZFLiu-XuEPJB17, ZFLiu-QiuAIP Adv.15, ZFLiu-HanAM14, ZFLiu-HuAM18}, thermal transparency~\cite{ZFLiu-XuPRA19a, ZFLiu-WangJAP18}, thermal coding~\cite{ZFLiu-ShangAPL18, ZFLiu-ZhouESEE19, ZFLiu-WangPRAP20, ZFLiu-YangPRAP23, ZFLiu-LeiIJHMT23, ZFLiu-HuAM19, ZFLiu-GuoAM22}, thermal chameleonlike metashells~\cite{ZFLiu-YangPRAP20, ZFLiu-XuSCPMA20, ZFLiu-YangEPL19-1, ZFLiu-XuPRA19, ZFLiu-XuEPL19, ZFLiu-XuEPJB19-2, ZFLiu-YangEPL19-2}, thermal sensor~\cite{ZFLiu-XuEPL20-1, ZFLiu-JinIJHMT20, ZFLiu-XuPRE19-1, ZFLiu-ShangJHT18, ZFLiu-WangJAP17}, and others~\cite{ZFLiu-DaiPR23, ZFLiu-ZhuangSCPMA22, ZFLiu-TianIJHMT21, ZFLiu-XuEPL20-2, ZFLiu-XuPRE19-2, ZFLiu-YangJAP19, ZFLiu-XuPRE18, ZFLiu-XuEPJB18, ZFLiu-JiCTP18, ZFLiu-ShangIJHMT18, ZFLiu-YangAPL17, ZFLiu-HuangPB17, ZFLiu-ShenAPL16, ZFLiu-Tan4, ZFLiu-QiuEur.Phys.J.Appl.Phys.15, ZFLiu-Tan15, ZFLiu-QiuIJHT14, ZFLiu-QiuEPL13}. However, the above works are all constrained to purely thermal conduction. Thermal convection, as another fundamental mechanism of heat transfer, is usually dominant in moving media such as fluids. After considering it, hybrid diffusive systems can revive many interesting phenomena in dissipative diffusion beyond pure conduction~\cite{ZFLiu-XuPANS23, ZFLiu-XuNSR23, ZFLiu-JinPNAS23, ZFLiu-YaoISci22, ZFLiu-LiPF22, ZFLiu-WangCPB22, ZFLiu-DaiPRAP22, ZFLiu-WangPRAP21, ZFLiu-WangATE21, ZFLiu-ZhangTSEP21, ZFLiu-XuEPL21-1, ZFLiu-WangICHMT20, ZFLiu-XuCPLEL20, ZFLiu-XuAPL20, ZFLiu-XuIJHMT20, ZFLiu-XuCPL20, ZFLiu-DaiPRE18, ZFLiu-DaiJAP18}. Nonreciprocal heat transfer can be easily implemented with the help of convection due to its directional nature, such as spatiotemporal modulation~\cite{ZFLiu-XuPRL22-1, ZFLiu-XuPRL22-2, ZFLiu-XuPRE21, ZFLiu-YangPRAP22, ZFLiu-TorrentPRL18, ZFLiu-CamachoNC20, ZFLiu-LiNC22} and angular momentum bias~\cite{ZFLiu-LiPRB21, ZFLiu-XuAPL21}. Transformation thermotics has been extended to thermal radiation~\cite{ZFLiu-XuESEE20, ZFLiu-XuPRAP20, ZFLiu-YangJAP20, ZFLiu-XuPRAP19}, nonlinear thermal conductivities~\cite{ZFLiu-LiPRL15, ZFLiu-ShenPRL16, ZFLiu-ZhuangPRE22, ZFLiu-DaiJNU21, ZFLiu-SuEPL20, ZFLiu-DaiIJHMT20, ZFLiu-WangPRE20, ZFLiu-YangPRE19, ZFLiu-DaiEPJB18, ZFLiu-Tan11}, and thermoelectrics~\cite{ZFLiu-LeiMTP23, ZFLiu-ZhuangIJMSD23, ZFLiu-QuEPL21, ZFLiu-LeiEPL21, ZFLiu-XuEPJB20, ZFLiu-WangPRA19, ZFLiu-QiuCTP14, ZFLiu-LiJAP10}. Besides, the concept of thermal metamaterials has also been generalized to other diffusion processes, such as particle diffusion~\cite{ZFLiu-XuPRE20, ZFLiu-ZhangATS22, ZFLiu-ZhangPRAP23} and plasma transport~\cite{ZFLiu-ZhangCPL22}. 

Recently, diffusion system has become a brand new platform to realize the topological phase and non-Hermitian physics. For heat conduction, there are two paradigms for implementing these phases. The first paradigm is spatially discretizing the diffusion equation~\cite{ZFLiu-YoshidaSP21}. Though with some imperfections in theory~\cite{ZFLiu-QiAM22}, the proposal has been experimentally confirmed to observe the edge state of one-dimensional (1D) Su-Schrieffer-Heeger model~\cite{ZFLiu-HuAM22}. As a natural generalization, this proposal can be extended to higher-dimensional systems, such as higher-order topological insulators~\cite{ZFLiu-LiuarXiv22-1, ZFLiu-QiarXiv23, ZFLiu-WuAM23}. The other paradigm is constructing the coupled ring chain structure. The structure consists of several rings vertically coupled in one direction to form a chain through interlayers. It has been proposed that the non-Hermitian skin effect~\cite{ZFLiu-CaoCP21, ZFLiu-CaoCPL22} and the quasicrystal~\cite{ZFLiu-LiuarXiv22-2} can be realized within this structure. Besides, the thermal analogue of coherent perfect absorption may support the establishment of non-Hermitian diffusive scattering theory~\cite{ZFLiu-LiNC22-2}.

However, when convection is introduced as a new degree of freedom into heat transfer, increasingly non-Hermitian phenomena and topological physics can be realized in thermal diffusion beyond purely conduction. In this review, we present some important works in this area, organized as follows. In Section ${\rm{\uppercase\expandafter{\romannumeral2}}}$, we will discuss the work about the realization of EP in diffusion systems. In addition, we will present how to implement high-order EPs and dynamically enclose EPs in thermal diffusion. In Section ${\rm{\uppercase\expandafter{\romannumeral3}}}$, we will discuss two works about the extensions of EP in convective thermal metamaterials, i.e., chiral thermal behaviour in the vicinity of EP and Weyl exceptional ring. In Section ${\rm{\uppercase\expandafter{\romannumeral4}}}$, we will introduce the topological physics in convection-diffusion systems. Two topological phases, a 1D topological insulator and a 2D quadrupole insulator, have been realized with the help of thermal convection. A concluding discussion is presented in Section ${\rm{\uppercase\expandafter{\romannumeral5}}}$, accompanying a reasonable outlook that may come true in the near future.

\section{${\rm{\textbf{\uppercase\expandafter{\romannumeral2}}}}$. Non-Hermitian physics in convective thermal metamaterials: The implementation of EP}

\begin{figure}[!ht]
\includegraphics[width=0.4\linewidth]{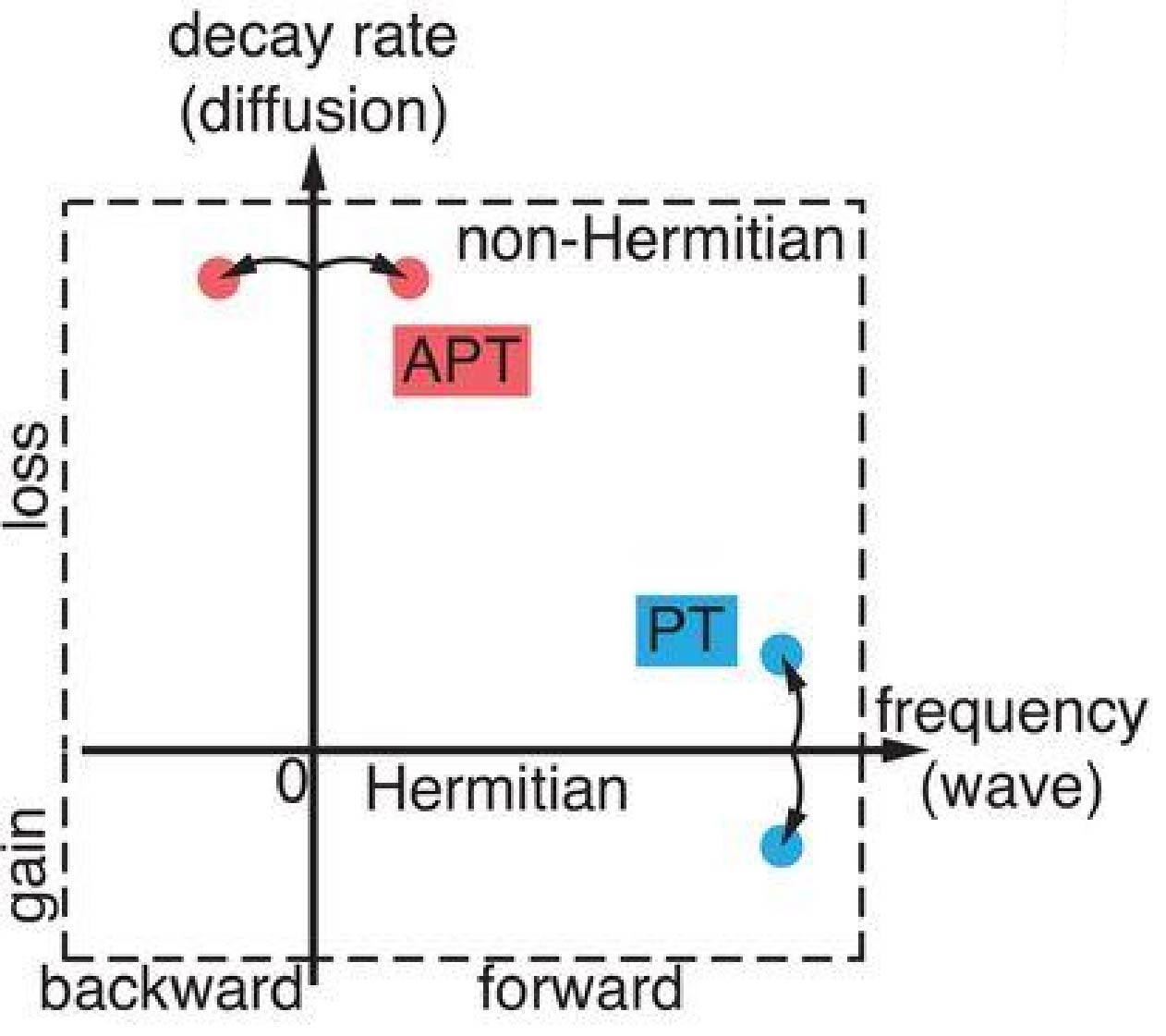}
\caption{Two paths to realize the PT-related phenomenon: the introduction of gain and loss in wave systems and the introduction of the forward and backward wave-like field in diffusion systems. Adapted from Ref.~\cite{ZFLiu-LiSci19}. With permission from the Author.}
\label{Fig1}
\end{figure}

The research on non-Hermitian convective thermal metamaterials starts from the implementation of anti-parity-time (APT) symmetry and EP in diffusion systems~\cite{ZFLiu-LiSci19}. This work lays the foundation of non-Hermitian convective thermal metamaterials because of the importance of EP in non-Hermitian physics. In the wave systems, realizing the PT symmetry is very intuitive by introducing balanced gain and loss [blue dots in Fig.~\ref{Fig1}]. However, it is not easy to realize the PT-related phenomenon in diffusion systems that are intrinsically dissipative (anti-Hermitian). This problem can be solved by introducing the forward and backward wave-like field, i.e., heat convection [red dots in Fig.~\ref{Fig1}], which acts as a Hermitian component. After the introduction of thermal convection, the diffusive system will become APT symmetric and its effective Hamiltonian anti-commutes with the PT operator. 

\begin{figure}[!ht]
\includegraphics[width=0.6\linewidth]{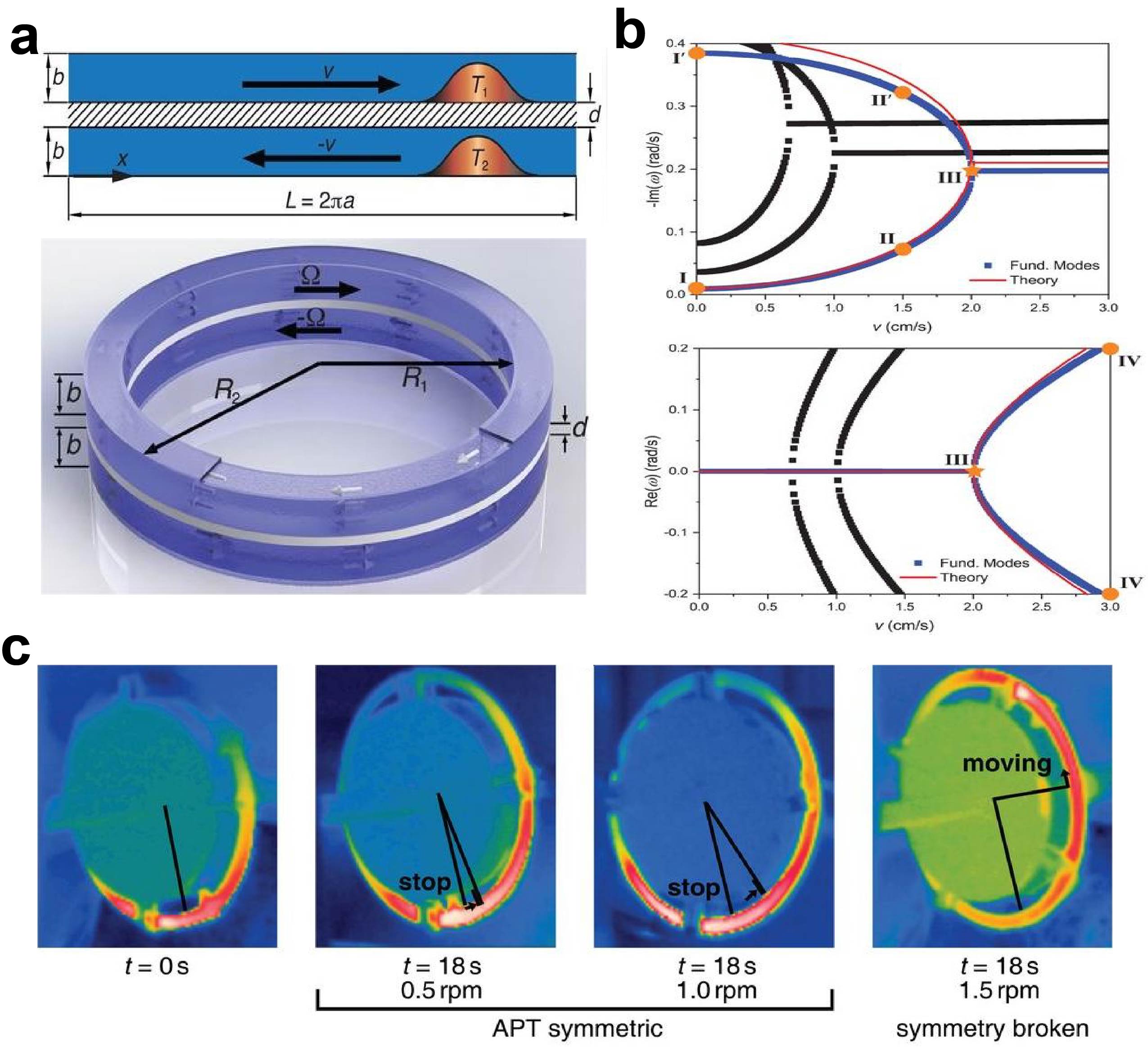}
\caption{APT symmetry and EP in convective thermal metamaterials. (a) The double ring model in the 2D plane (upper panel) and its 3D diagram (lower panel). (b) Decay rates and eigenfrequencies of double ring model. The blue squares are the simulation results for the fundamental mode, and the red lines are the theoretical eigenvalues of the effective Hamiltonian. (c) Infrared images of the temperature profiles on ring 1 at $t$ = 0 s and $t$ = 18 s with rotating velocity 0.5 rpm, 1.0 rpm, and 1.5 rpm. Adapted from Ref.~\cite{ZFLiu-LiSci19}. With permission from the Author.}
\label{Fig2}
\end{figure}

The basic structure is two coupled rings with equal but opposite rotating velocities [see the schematic diagram in Fig.~\ref{Fig2}(a)]. The governing heat transfer equations for this double ring model can be written as 
\begin{align}
\frac{{\partial}T_{1}}{{\partial}t}=D\frac{{\partial}^{2}T_{1}}{{\partial}x^2}-v\frac{{\partial}T_{1}}{{\partial}x}+h(T_2-T_1) \\
\frac{{\partial}T_{2}}{{\partial}t}=D\frac{{\partial}^{2}T_{2}}{{\partial}x^2}+v\frac{{\partial}T_{2}}{{\partial}x}+h(T_1-T_2)
\label{double_ring_conv_eq}
\end{align}
where $D$ is the diffusivity of rings, $h$ is the heat exchange rate between two rings, and $v$ is the rotating velocity of rings. Using the plane-wave solution, the effective APT symmetric Hamiltonian is obtained as:
\begin{equation}
{\hat{H}}=\left(\begin{matrix}
            -i(k^{2}D+h)+kv & ih
            \\
            ih & -i(k^{2}D+h)-kv
            \end{matrix}\right)
\label{double_ring_conv_Ham}
\end{equation}  
where $k$ is the wave number. The corresponding eigenvalues of the effective Hamiltonian are 
\begin{align}
\omega_{\pm}=-i\left[(k^{2}D+h){\pm}\sqrt{h^{2}-k^{2}v^{2}}\right]
\label{double_ring_conv_eigen}
\end{align}
Then the EP emerges when $h^{2}=k^{2}v_{\rm{EP}}^2$. The fundamental mode $k=1/R$ is chosen because only the slowest decaying mode can be observed. The theoretical decay rates and eigenfrequencies of the effective Hamiltonian are shown in Fig.~\ref{Fig2}(b), which are in accordance with the simulation results. The two branches of decay rates coalesce at the EP, and the nonzero eigenfrequency appears as the $v$ increases. In the APT unbroken phase, the temperature profile of ring 1 remains motionless, and the maximum temperature point is fixed after evolving a certain time [see the middle two subfigures in Fig.~\ref{Fig2}(c)]. However, the temperature profile keeps moving because of nonzero eigenfrequency in the APT broken phase [see the right subfigure in Fig.~\ref{Fig2}(c)]. The key point behind these thermal behaviours is the competition between thermal convection and coupled thermal conduction.

\begin{figure}[!ht]
\includegraphics[width=0.6\linewidth]{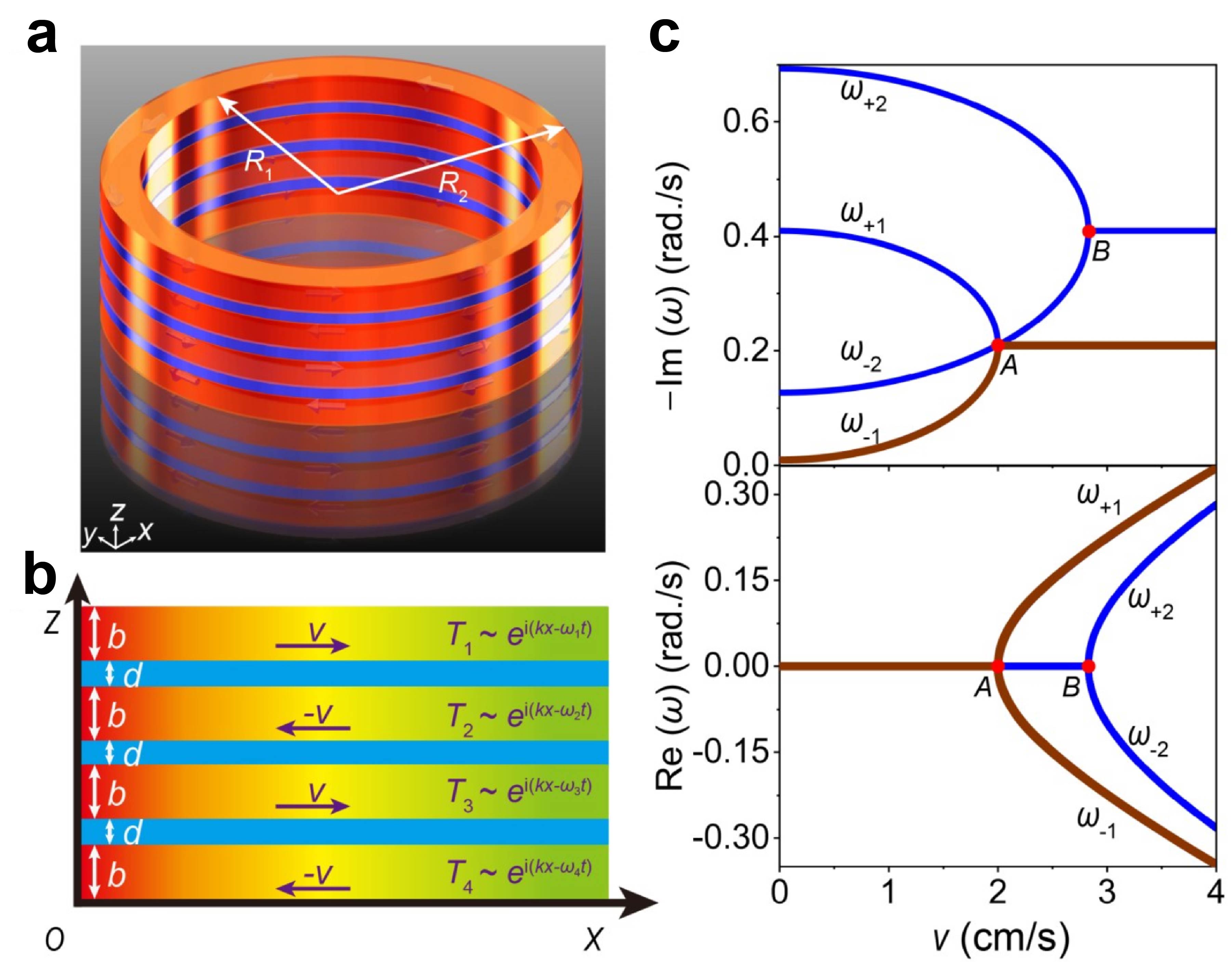}
\caption{High-order EP in convective thermal metamaterials. Schematic diagram of (a) 3D quadruple ring model and (b) its corresponding 2D model. (c) The imaginary part and real part of eigenvalues vs. the background flow velocity. Red dots at $A$ and $B$ mark the EPs. Adapted from Ref.~\cite{ZFLiu-CaoESEE20}. With permission from the Author.}
\label{Fig3}
\end{figure}

Further increasing the number of rings, high-order EP can be achieved in convection-diffusion systems, which is robust against perturbations and phase oscillations~\cite{ZFLiu-CaoESEE20}. The model is the quadruple ring structure with opposite but equal velocities between adjacent channels [see in Fig.~\ref{Fig3}(a) and (b)]. Similar with the derivation in Ref.~\cite{ZFLiu-LiSci19}, the effective Hamiltonian is a four by four matrix:
\begin{equation}
{\hat{H}}=\left(\begin{matrix}
            -i(k^{2}D+h)-kv & ih & 0 & 0
            \\
            ih & -i(k^{2}D+h)-ih+kv & ih & 0
            \\
            0 & ih & -i(k^{2}D+h)-ih-kv & ih 
            \\
            0 & 0 & ih & -i(k^{2}D+h)+kv 
            \end{matrix}\right)
\label{quatre_ring_conv_Ham}
\end{equation}  
After diagonalizing the Hamiltonian, a third-order EP can be obtained at the critical velocity when APT phase transition occurs [point $A$ in Fig.~\ref{Fig3}(c)]. In the APT symmetric phase, the temperature profile remains stationary with a phase-locked difference despite the velocity perturbation. While in the APT broken phase, the temperature field starts to dynamically evolve with phase oscillations regardless of the initial conditions. 

\begin{figure}[!ht]
\includegraphics[width=0.6\linewidth]{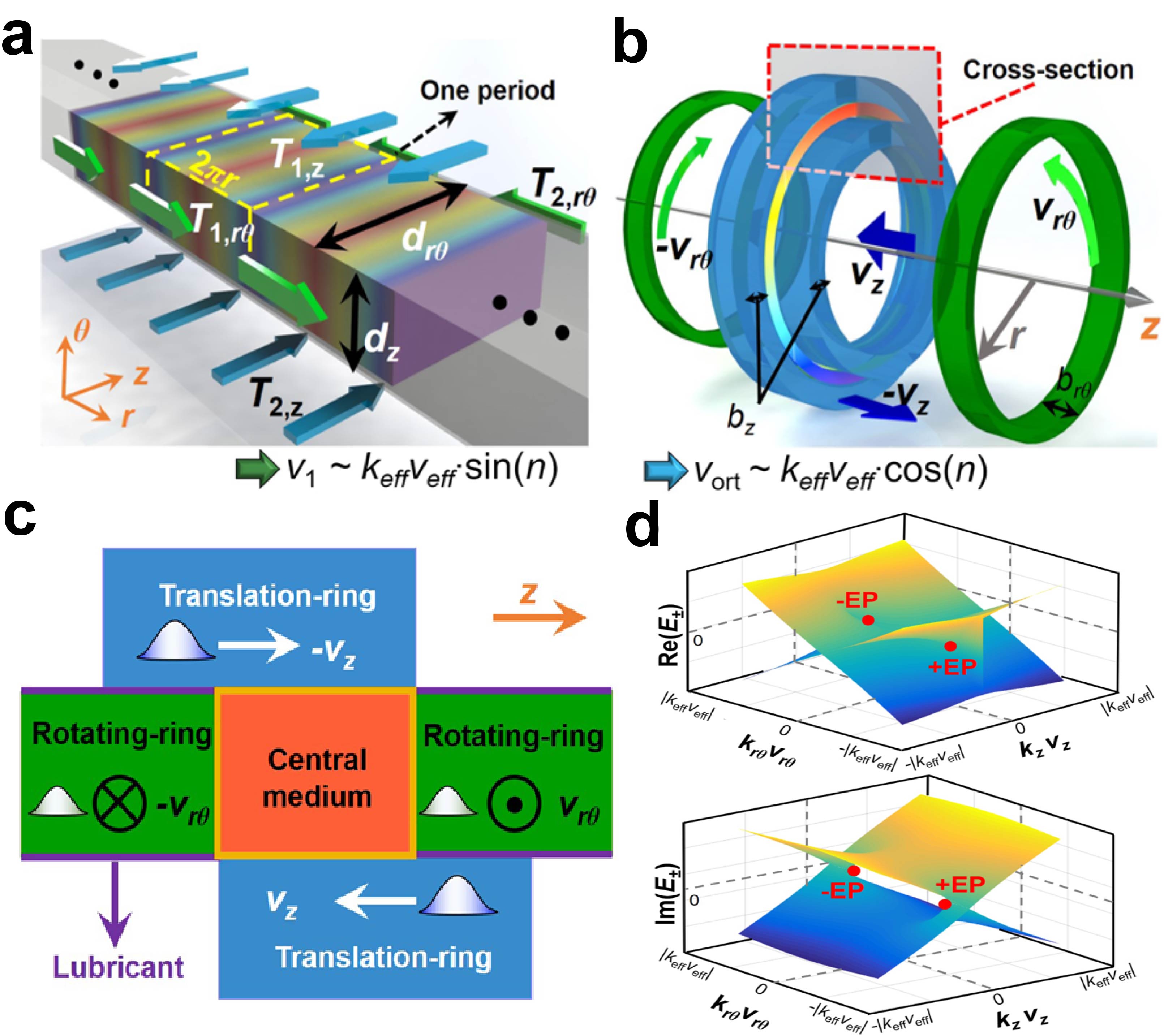}
\caption{Dynamic encircling around the EP in convective thermal metamaterials. (a) Schematic of the periodic convection-conduction system with two orthogonal pairs of advections. (b) Schematic diagram of the multiple ring system. (c) The cross section view labelled in the red dashed border of Fig.~\ref{Fig4}(b). (d) Real part and imaginary part of the eigenvalues of the effective Hamiltonian. Adapted from Ref.~\cite{ZFLiu-XuPRL21}. With permission from the Author.}
\label{Fig4}
\end{figure}

In order to reveal the nontrivial topology of EPs in non-Hermitian physics, a sufficiently large parameter space is essential for dynamical EP encircling. However, the fixed and positive conductivity and structural parameters prevent the encircling from occurring in thermal diffusion. To address this problem, a recent work has created an orthogonal convection space to provide enough degrees of freedom~\cite{ZFLiu-XuPRL21}. The first step is to mimic two pairs of orthogonal oscillations as a complex parameter space [see Fig.~\ref{Fig4}(a)]. In order to easily implement such periodic convection, it is necessary to devise a multiple ring system consisting of two orthogonal pairs of counter-convective rings. One pair is rotating in $r-{\theta}$ space, and the other pair is translating along the $z$ direction [see the 3D diagram in Fig.~\ref{Fig4}(b) and the corresponding cross-section view in Fig.~\ref{Fig4}(c)]. Substituting the wave-like solution into the heat transfer equation, the effective Hamiltonian can be written as 
\begin{equation}
\hat{H}=i\left(\begin{matrix}
         -\left(\dfrac{\kappa}{{\rho}c}k_{r{\theta}}^2+2m\right)-k_{\rm{eff}}v_{\rm{eff}}\left[{\rm{cos}}(n)+i{\rm{sin}}(n)\right] & 2m
         \\
         2m & -\left(\dfrac{\kappa}{{\rho}c}k_{r{\theta}}^2+2m\right)+k_{\rm{eff}}v_{\rm{eff}}\left[{\rm{cos}}(n)+i{\rm{sin}}(n)\right]  
\end{matrix}\right)
\label{multiple_ring_Ham}
\end{equation}  
where $\kappa$, $\rho$, and $c$ are the thermal conductivity, mass density, and heat capacity of rings. $m$ is the heat exchange rate, which is fixed and positive. For clarity, the orthogonal convection space is denoted as $\left[k_{\rm{eff}}v_{\rm{eff}}{\rm{cos}}(n), k_{\rm{eff}}v_{\rm{eff}}{\rm{sin}}(n)\right]$, where $k_{\rm{eff}}v_{\rm{eff}}=\sqrt{(k_{r{\theta}}v_{r{\theta}})^2+(k_{z}v_{z})^2}$ and $n={\rm{arctan}}(k_{r{\theta}}v_{r{\theta}}/k_{z}v_{z})$. $v_{r{\theta}}$ and $v_{z}$ are the rotating and translating velocities of rings. The effective wave numbers are $k_{r{\theta}}=1/r$ in $r-{\theta}$ plane and $k_{z}=1/d$ along the $z$ direction, where $r$ is the radius of ring and $d$ is the thickness of the medium layer. The complex spectrum of this effective Hamiltonian is calculated in Fig.~\ref{Fig4}(d), where two EPs can be found. 

\begin{figure}[!ht]
\includegraphics[width=0.7\linewidth]{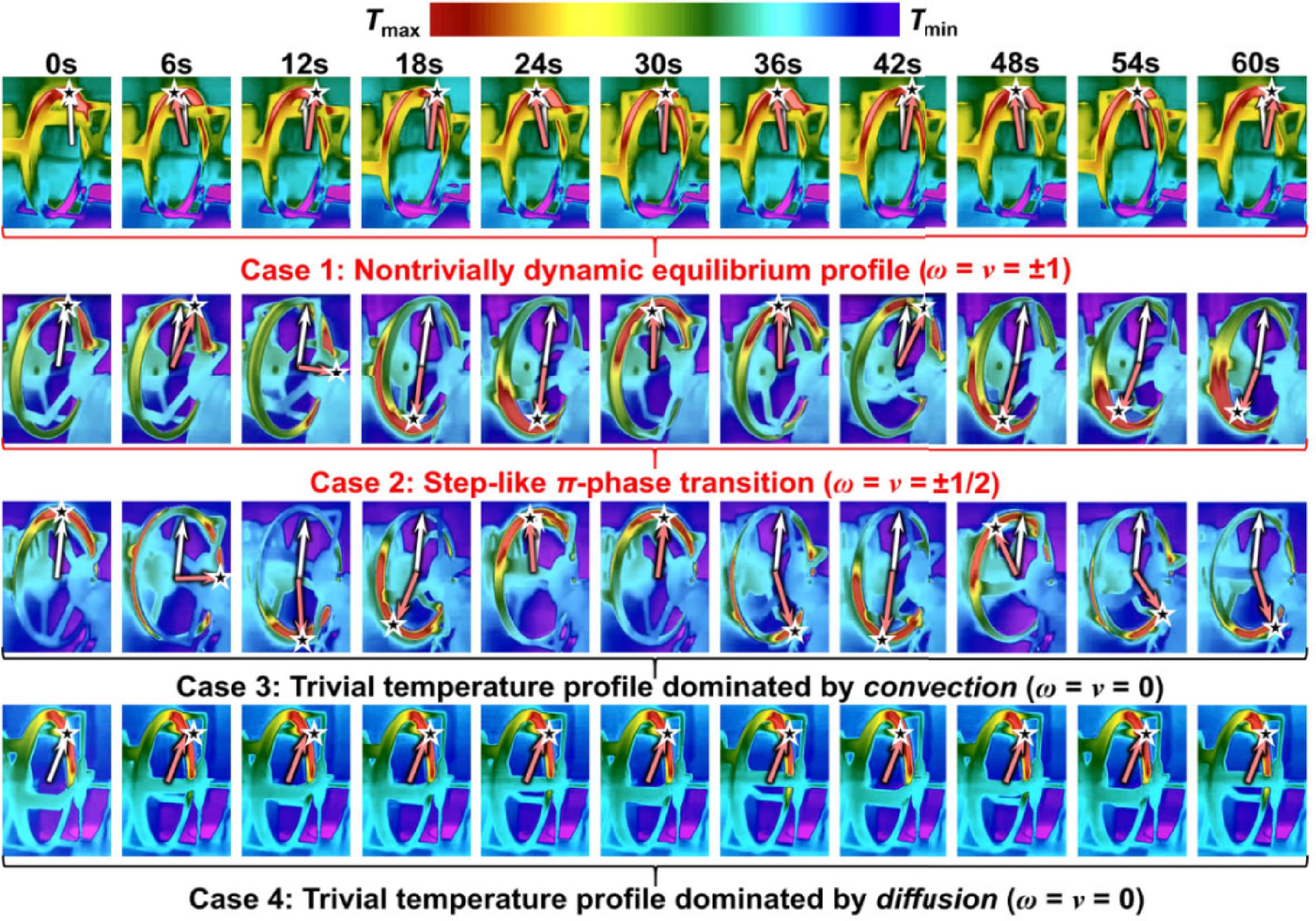}
\caption{Measured temperature profiles of four cases on the surface of the central medium. The black stars indicate the locations of the maximum temperature points. The white and red arrows indicate the initial and measured azimuthal angles. $\omega$ and $\nu$ are the eigenstate winding number and eigenvalue vorticity of EP. Adapted from Ref.~\cite{ZFLiu-XuPRL21}. With permission from the Author.}
\label{Fig5}
\end{figure}

The topological properties of the EP are determined by the geometric phase generated by a dynamically encircling trajectory. It is related to two quantized topological invariants: the eigenstate winding number $\omega$~\cite{ZFLiu-LeykamPRL17} and the eigenvalue vorticity $\nu$~\cite{ZFLiu-ShenPRL18}. When the evolutionary trajectory of convections encircles two EPs, both topological invariants are integer numbers, revealing a non-Hermitian thermal topology. The temperature profile of the surface of the central medium exhibits a dynamic-equilibrium distribution without much deviation, and so does the location of the maximum temperature point [see the top subfigure in Fig.~\ref{Fig5})]. In this case, the initial state undergoes a full Riemann sheet during the adiabatic evolution and returns to itself at the end of one period. By comparison, when the trajectory encloses only one EP, both topological invariants are half-integer. The temperature profile and the location of maximum temperature show a step-like $\pi$-phase transition after one period [see the second top subfigure in Fig.~\ref{Fig5}. The eigenstate will exchange to another state with a $\pi$-phase winding after one period and return to its initial position after two periods. Furthermore, both topological invariants are zero when the EPs are not enclosed in the locus. Then the system becomes a conventional thermal metamaterial. For large advection modulations, the maximum temperature locations continuously change with the convection [see the third top subfigure in Fig.~\ref{Fig5}. While for small convection, the maximum temperature locations keep unchanged [see the fourth top subfigure in Fig.~\ref{Fig5}. In this way, the dynamical encircling of the EP in thermal diffusion is realized and the configurable topological phase transition of thermal systems is observed for the first time. 

Meanwhile, a separate work shows the geometric phase induced by heat convection in diffusion systems~\cite{ZFLiu-XuIJHMT21}. Based on the double ring model, the temperature field will accumulate an extra $\pi$ phase when the cyclic evolution of time-varying rotating velocity contains the EP.

\section{${\rm{\textbf{\uppercase\expandafter{\romannumeral3}}}}$. Non-Hermitian physics in convective thermal metamaterials: The extension of EP}

In this section, we will introduce two extended researches based on the previous works about EP in thermal diffusion, namely the chiral behaviour in the vicinity of EP~\cite{ZFLiu-XuPRL23} and the Weyl exceptional ring~\cite{ZFLiu-XuPNAS22}. 

\begin{figure}[!ht]
\includegraphics[width=0.6\linewidth]{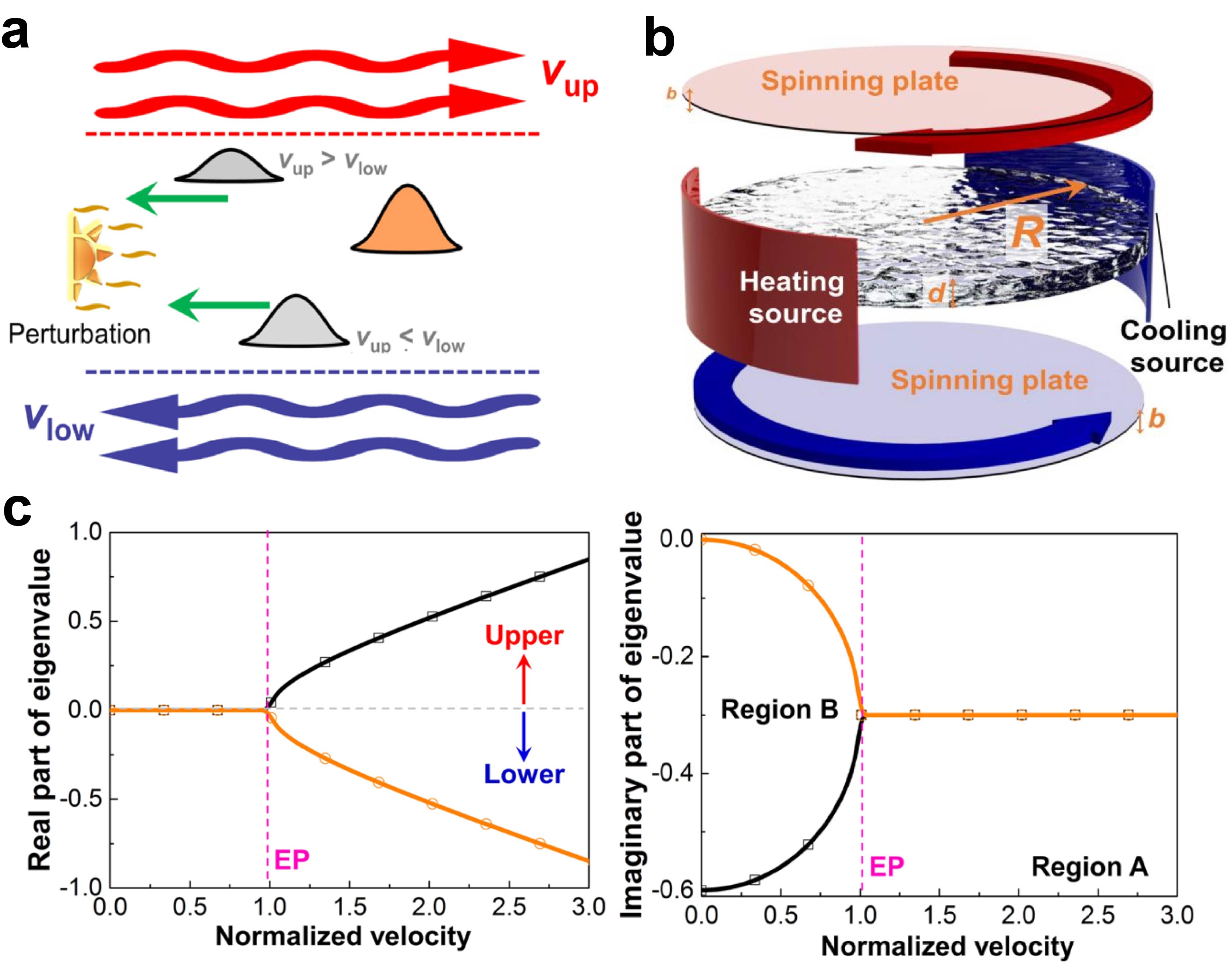}
\caption{Chiral behaviour in convective thermal metamaterials. (a) Schematic diagram of the chirality in heat transport under constant thermal perturbations and advections. Two wave-like advections are imposed on the upper and lower boundaries. The eigenstates (gray wave packets) exhibit the same directions (green arrows). (b) Schematic experimental setup for studying the chiral heat transport in a non-Hermitian thermal system with counter-spinning plates. (c) Real and imaginary part of eigenvalues of the effective Hamiltonian. The normalized velocity is $|v_{\rm{up}}|+|v_{\rm{low}}|$. Region A (B) denotes the APT broken (unbroken) phase. Adapted from Ref.~\cite{ZFLiu-XuPRL23}. With permission from the Author.}
\label{Fig6}
\end{figure}

The chiral behaviour at the EP has been proposed to achieve asymmetric wave propagation and unidirectional invisibility in carefully designed whispering-gallery-mode resonators~\cite{ZFLiu-WangNP20, ZFLiu-PengPNAS16, ZFLiu-PengNP14}. However, the chirality has not been demonstrated in thermal diffusion so far. The reason is that the moving temperature field only follows the dominating advection introduced into the system~\cite{ZFLiu-LiSci19, ZFLiu-XuPRL21, ZFLiu-XuNP22, ZFLiu-XuPNAS22}. A recent work has shown the chiral heat transport after imposing the thermal perturbation~\cite{ZFLiu-XuPRL23}. The asymmetric thermal profile can be observed regardless of the directions and magnitudes of advections [see the schematic diagram in Fig.~\ref{Fig6}(a)]. The proposed experimental fluid system is shown in Fig.~\ref{Fig6}(b). Two thin plates are employed on the central fluid to impose the counter-advection at its upper and lower boundaries. The heating and cooling sources are directly imposed as perturbations on the boundary of the circular fluid region. These sources are retained after observing the steady-state thermal distribution, and advection is subsequently activated. After substituting the initial perturbed temperature field into the heat transfer equation, the effective Hamiltonian for the global system can be written as:
\begin{equation}
\hat{H}=i\left(\begin{matrix}
         -\left(\dfrac{k^{2}\kappa_{0}}{\rho_{0}c_{0}}+\dfrac{h}{\rho_{0}c_{0}b}\right)-ik|v_{\rm{up}}| & \dfrac{k^{2}\kappa_{0}}{\rho_{0}c_{0}}+\dfrac{h}{\rho_{0}c_{0}b}
         \\
         -\dfrac{k^{2}\kappa_{0}}{\rho_{0}c_{0}}+\dfrac{h}{\rho_{0}c_{0}b} & -\left(\dfrac{k^{2}\kappa_{0}}{\rho_{0}c_{0}}+\dfrac{h}{\rho_{0}c_{0}b}\right)+ik|v_{\rm{low}}|
\end{matrix}\right)
\label{chiral_Ham}
\end{equation}  
where $k$, $h$, and $b$ denote the effective wave number, the convective heat transfer coefficient, and the thickness of the spinning plate. $v_{\rm{up}}$ and $v_{\rm{low}}$ are the rotating velocities of the upper and lower plates. An EP can be found in the spectrum of this Hamiltonian [see in Fig.~\ref{Fig6}(c)]. 

\begin{figure}[!ht]
\includegraphics[width=\linewidth]{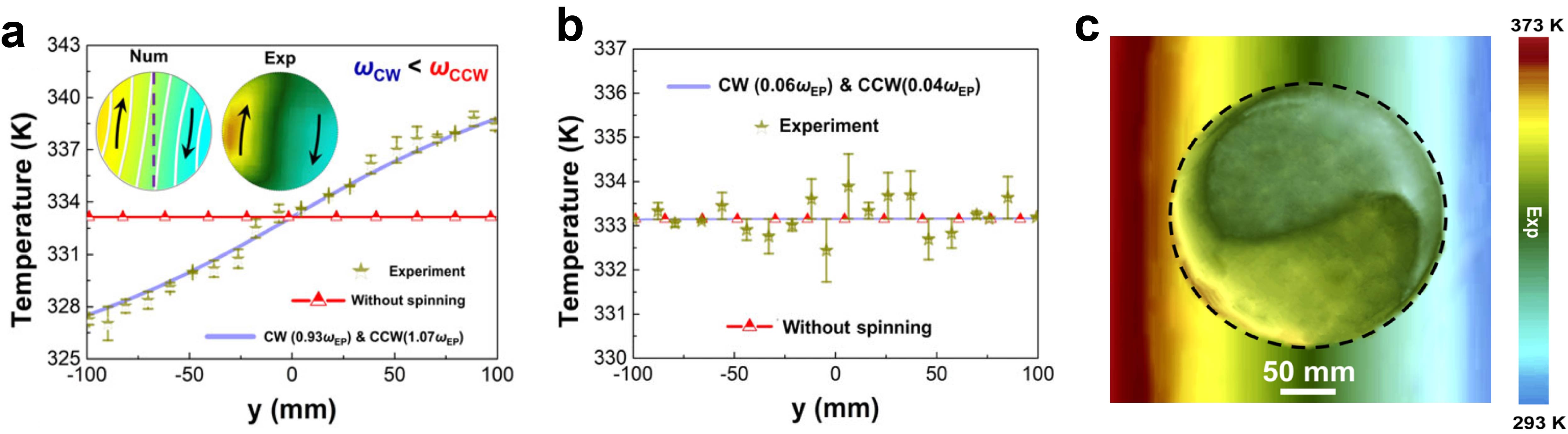}
\caption{Temperature field of chiral and nonchiral heat transport. (a) The temperature distribution in region A on the measured line (purple dashed line in the inset). The insets present the calculated (left) and experimental (right) thermal distributions. (b) The temperature distribution in region B on the measured line (the same location as in the inset of Fig.~\ref{Fig6}(a)). (c) The unidirectional thermal twisting profile induced by the chiral and nonchiral heat transport. Adapted from Ref.~\cite{ZFLiu-XuPRL23}. With permission from the Author.}
\label{Fig7}
\end{figure}

Then the temperature field of the central fluid is explored in different regions. In the vicinity of EP in region A (APT broken phase), the thermal field at the fluid interface moves towards the CW direction, though the rotating velocity at CW is smaller than the one at CCW [see in Fig.~\ref{Fig7}(a)]. These positive phase deflections are caused by the velocity difference of the imposed advection. However, when the advection is extremely large, that is, away from the EP in region A, the temperature field becomes homogeneous without any chiral behaviour. When in region B (APT symmetric phase), the temperature field is unbiased with nonchiral heat transport [see Fig.~\ref{Fig7}(b)]. Moreover, the chirality vanishes at the EP due to purely imaginary eigenvalues and zero phase deflections in the eigenstates. Chiral thermal transport refers to an effectively anisotropic thermal conductivity, whereas the nonchiral one can be treated as isotropic. Based on these facts, the combination of nonchiral and chiral heat transports within one fluid may attribute to a twisted thermal profile in Fig.~\ref{Fig7}(c). Thus, the chiral temperature field exhibits a large power in the free manipulation of the heat current.  

Weyl point is a band degeneracy point whose dispersion satisfies the Weyl equation in high-energy physics~\cite{ZFLiu-WanPRB11, ZFLiu-ArmitageRMP18}. It can be regarded as a magnetic monopole in 3D momentum space and possesses a quantized Chern number. Moreover, due to the fermion doubling theorem, the Weyl points must pair with opposite topological charges. When the non-Hermitian term is introduced into Weyl Hamiltonian, the Weyl point will evolve into the Weyl exceptional ring composed of EPs~\cite{ZFLiu-XuPRL17}. The Weyl exceptional ring has been implemented in topological photonics~\cite{ZFLiu-CerjanNP19} and topological acoustics~\cite{ZFLiu-LiuPRL22}. 

\begin{figure}[!ht]
\includegraphics[width=0.6\linewidth]{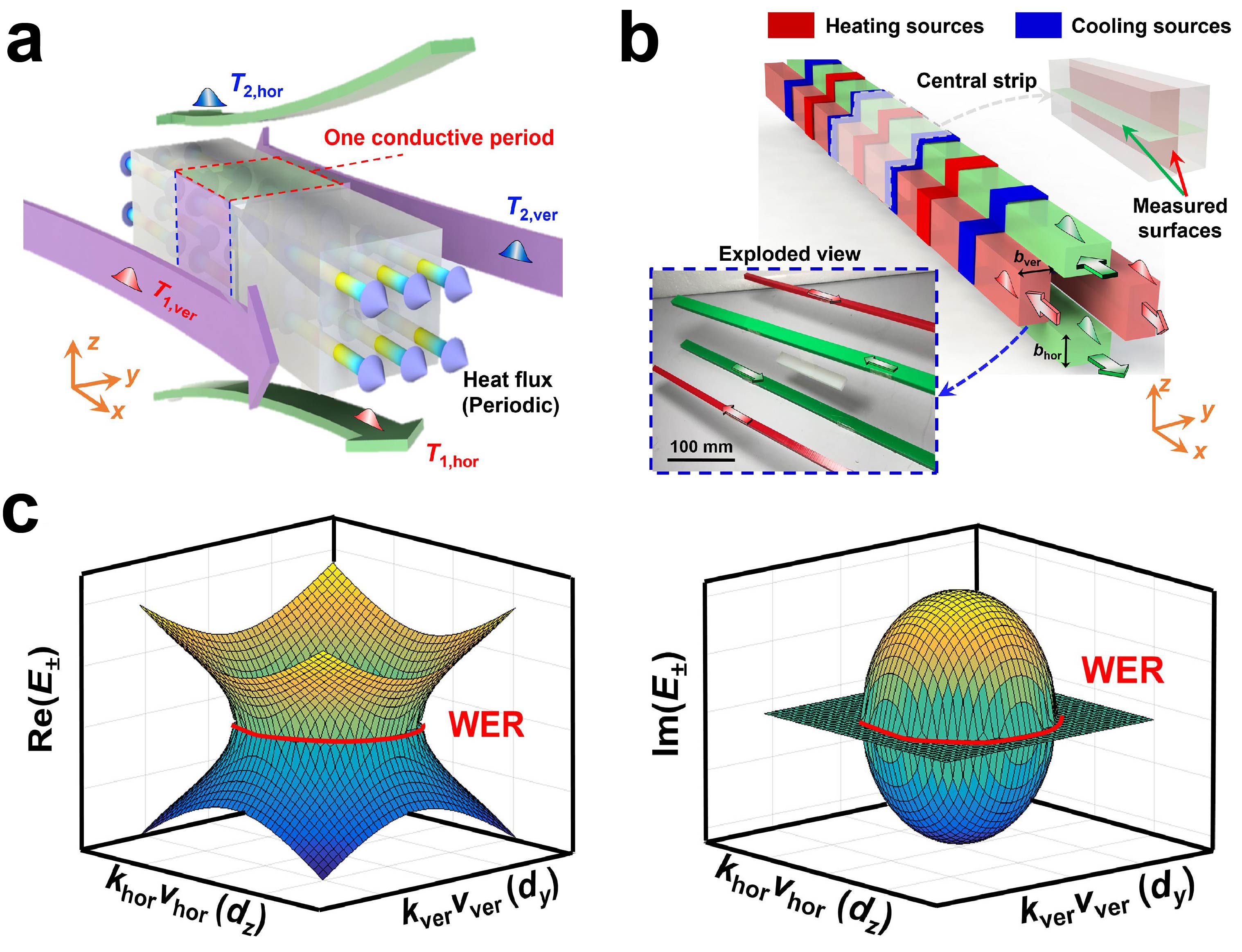}
\caption{Weyl exceptional ring in convective thermal metamaterials. (a) Schematic diagram of diffusive Weyl exceptional ring model with four counter convections in orthogonal surfaces. The green arrows indicate the horizontal convections imposed on the upper and bottom surfaces. The purple arrows denote the vertical convections imposed at the lateral surfaces. (b) Schematic diagram of experimental setup which exhibits the Weyl exceptional ring. The heating and cooling sources are imposed periodically and alternatively on the entire system. The green and red strips indicate horizontal and vertical convection. The arrows denote the initial direction of motion of these strips. The upper and lower insets are enlarged views of the central strip and its decompositions. (c) Real and imaginary part of the eigenvalues of diffusive Weyl exceptional ring Hamiltonian. Adapted from Ref.~\cite{ZFLiu-XuPNAS22}. With permission from the Author.}
\label{Fig8}
\end{figure}

A recent work demonstrates the realization of Weyl exceptional ring in diffusion systems~\cite{ZFLiu-XuPNAS22}. Inspired by helical waveguides with helicoidal properties in optics~\cite{ZFLiu-CerjanNP19}, a central strip is surrounded by four counter-convection components in orthogonal spaces [see the schematic diagram in Fig.~\ref{Fig8}(a)]. To realize the periodic wave-like field, the initial temperature profile on the combined structure is imposed periodically and alternatively by the heating and cooling sources [see the experimental setup in Fig.~\ref{Fig8}(b)]. Substituting the periodic wave-like solution into a heat transfer process, the diffusive Weyl exceptional ring Hamiltonian is:          
\begin{equation}
\hat{H}=\left(\begin{matrix}
         -i\left[\dfrac{\kappa}{{\rho}c}(k_{\rm{hor}}^2+ik_{\rm{ver}}^2)+m_{\rm{hor}}+im_{\rm{ver}}\right]+k_{\rm{hor}}v_{\rm{hor}} & i(m_{\rm{hor}}+im_{\rm{ver}}+k_{\rm{ver}}v_{\rm{ver}})
         \\
         i(m_{\rm{hor}}+im_{\rm{ver}}-k_{\rm{ver}}v_{\rm{ver}}) & -i\left[\dfrac{\kappa}{{\rho}c}(k_{\rm{hor}}^2+ik_{\rm{ver}}^2)+m_{\rm{hor}}+im_{\rm{ver}}\right]-k_{\rm{hor}}v_{\rm{hor}}
\end{matrix}\right)
\label{WER_Ham}
\end{equation}  
where $\kappa$, $\rho$, and $c$ are the central strip's thermal conductivity, mass density, and heat capacity. $v_{\rm{hor}}$ and $v_{\rm{ver}}$ are the velocities of horizontal and vertical strips. The effective wave numbers can be indicated as $k_{\rm{hor}}=k_{\rm{ver}}=1/L$ where $L$ is the length of the central strip in one period. $m_{\rm{hor}}$ and $m_{\rm{ver}}$ denote the heat exchange rate in the horizontal and vertical plane. Because of the large vertical thickness of the central strip and material parameters of vertical strips, $m_{\rm{ver}}$ is proportional to the square root of $v_{\rm{ver}}$. So a 3D synthetic parameter space $\left(m_{\rm{ver}}, k_{\rm{ver}}v_{\rm{ver}}, k_{\rm{hor}}v_{\rm{hor}}\right)$ can be generated, and it provides enough degrees of freedom to encircle the Weyl exceptional ring. When $m_{\rm{ver}}$ keeps unchanged, a Weyl exceptional ring can be found in the spectrum as shown in Fig.~\ref{Fig8}(c). 

\begin{figure}[!ht]
\includegraphics[width=0.8\linewidth]{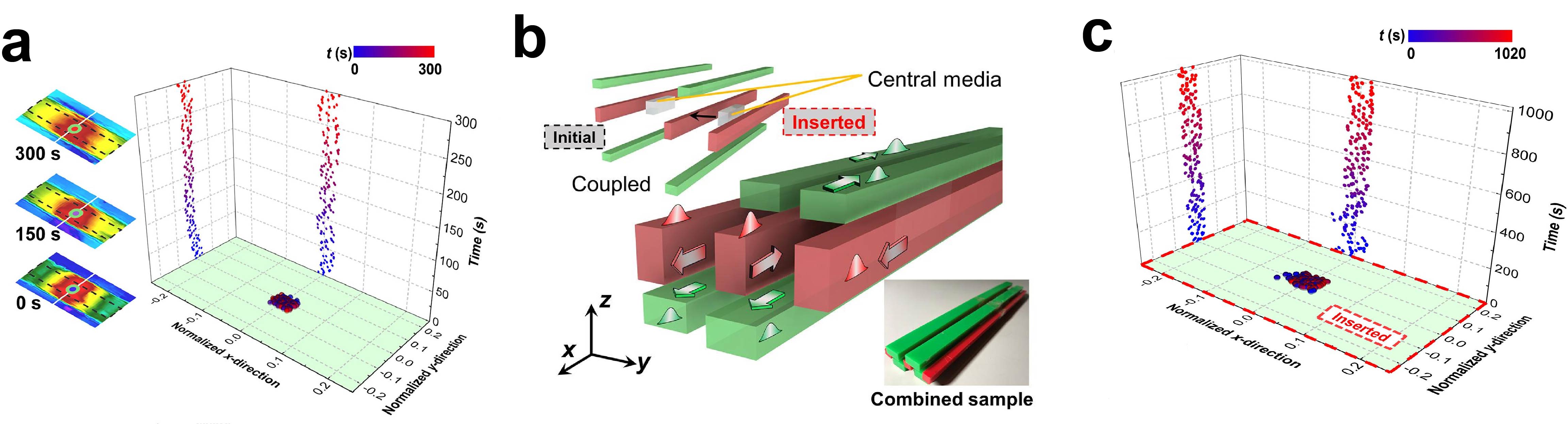}
\caption{Temperature field of Weyl exceptional ring and Fermi arc. (a) The experimental evolutions of maximum temperature point in the horizontal plane when the integration surface encloses the Weyl exceptional ring. The temperature profiles at specific moments are shown in the side. (b) Schematic diagram of the experimental setup with an inserted system that shows the Fermi arc state. (c) The experimental maximum temperature evolution in the horizontal plane of the inserted system for the Fermi arc. Adapted from Ref.~\cite{ZFLiu-XuPNAS22}. With permission from the Author.}
\label{Fig9}
\end{figure}

Next the thermal behaviour of Weyl exceptional ring is investigated. If the integration surface encircles the whole Weyl exceptional ring, the temperature profile and the location of the maximum temperature point remain stationary [see Fig.~\ref{Fig9}(a)], which is a signature of nontrivial topology. Besides, the Fermi arc state still exists but is suppressed in the Weyl exceptional ring system~\cite{ZFLiu-XuPRL17}. To observe the Fermi arc in diffusion systems, an additional subsystem should be inserted into the initial system, and it contributes to an opposite topological charge with an opposite integral direction [see the schematic diagram in Fig.~\ref{Fig9}(b)]. The evolution of maximum temperature positions for the inserted system also remains stationary with robustness [see in Fig.~\ref{Fig9}(c)], and so does one for the original system.

\section{${\rm{\textbf{\uppercase\expandafter{\romannumeral4}}}}$. Topological phenomena in convective thermal metamaterials}

A problem that limits the development of thermal topology is that only the temperature field of the slowest decaying branch can be observed in purely diffusive systems. But the introduction of thermal convection can evade this disadvantage. Essentially, thermal convection is a wave-like field that induces a wave topology rather than a diffusive one. Based on this fact, many interesting and important topological states can be realized in diffusion systems. In this section, we will introduce two works about topological physics in convective thermal metamaterials, i.e., 1D topological insulator~\cite{ZFLiu-XuNP22} and 2D quadrupole topological insulator~\cite{ZFLiu-XuNC23}.

\begin{figure}[!ht]
\includegraphics[width=0.6\linewidth]{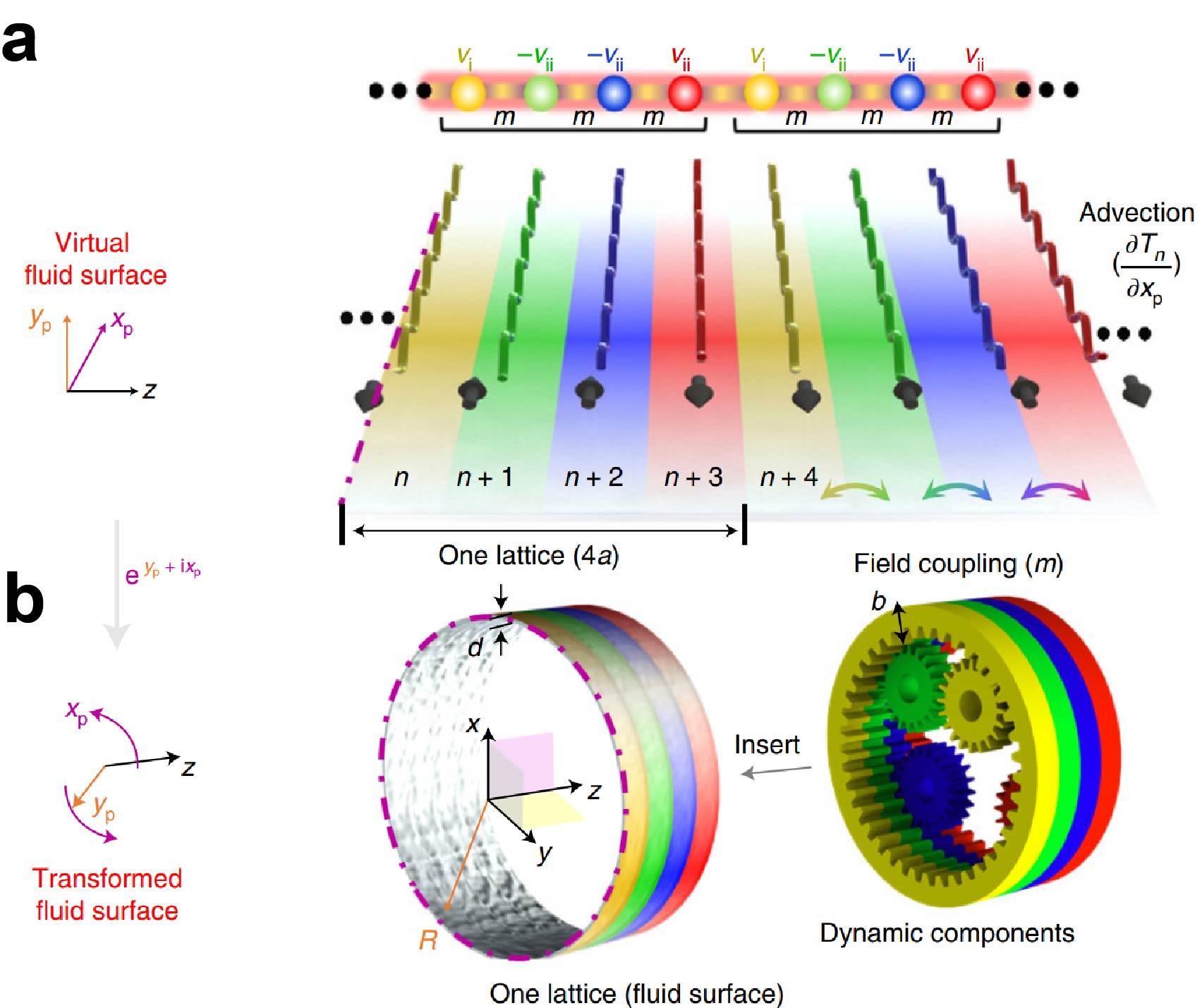}
\caption{Coupled ring chain structure. (a) Schematic diagram of a planar fluid surface with periodic convections in the virtual space ($x_{p}, y_{p}, z$). These periodic convections are imposed along the $x_{p}$ direction and arranged along the $z$ direction to form a lattice with four units. (b) Schematic diagram of four units for coupled ring chain structure in the transformed space ($x, y, z$). Adapted from Ref.~\cite{ZFLiu-XuNP22}. With permission from the Author.}
\label{Fig10}
\end{figure}

The topological edge state can be probed by tracing the temperature evolution in pure conduction systems~\cite{ZFLiu-YoshidaSP21, ZFLiu-HuAM22}. However, the temperature field for the edge states decays rapidly and is not clearly observed, which is unfavorable for localized thermal management. With the help of convection modulation, a recent work states that the edge state for a 1D system has been observed with a robust temperature profile~\cite{ZFLiu-XuNP22}. A planar fluid surface is considered in the virtual space ($x_{p}, y_{p}, z$) [see in Fig.~\ref{Fig10}(a)]. The periodic heat convections along the $x_{p}$ direction are imposed on the corresponding regions marked by different colours. One lattice with four units along the $z$ direction has been generated. Furthermore, the two ends for one unit along the $x_{p}$ direction should be connected to form a ring in the transformed space ($x, y, z$) for practical implementation [see the ring structure in Fig.~\ref{Fig10}(b)]. Substituting the wave-like solution into the heat transfer process, the Bloch Hamiltonian for this coupled ring chain structure can be written as 
\begin{equation}
\hat{H}=i\left(\begin{matrix}
ikv_{\rm{\romannumeral1}} & m & 0 & me^{ik_{t}4a}
\\
m & -ikv_{\rm{\romannumeral2}} & m & 0
\\
0 & m & -ikv_{\rm{\romannumeral1}} & m
\\
me^{-ik_{t}4a} & 0 & m & ikv_{\rm{\romannumeral2}}      
\end{matrix}\right)-i\left(\frac{\kappa}{{\rho}c}k^2+m\right)I_{4{\times}4}
\label{1D_edge_Ham}
\end{equation}  
where $\kappa$, $\rho$, and $c$ are the thermal conductivity, mass density, and heat capacity of rings. $m=\kappa/{{\rho}cbd}$ is the heat exchange rate between two adjacent rings. $b$ and $d$ are the thickness of the ring and interlayer. $v_{\rm{\romannumeral1}}$ and $v_{\rm{\romannumeral2}}$ are the modulation velocities and $a$ is the length of one unit. $k$ denotes the fundamental wave number of ring and $k_{t}$ indicates the effective Bloch wave number. $I$ is the identity matrix. The topological properties of this system are nontrivial with a pair of edge states under this convective arrangement. 

\begin{figure}[!ht]
\includegraphics[width=0.8\linewidth]{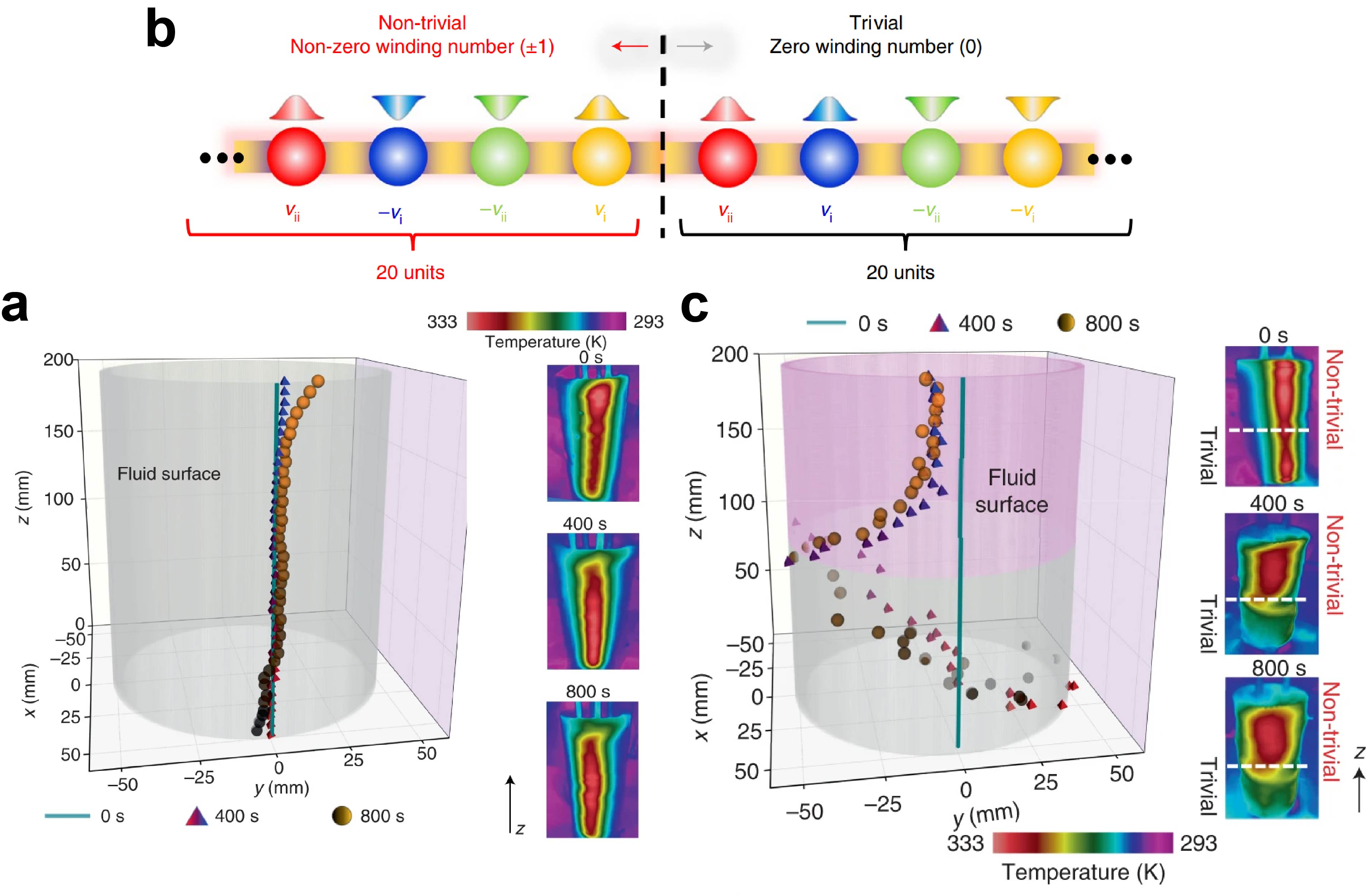}
\caption{Temperature field of topological edge state and interface state. (a) The experimental locations of the maximum temperature point for each unit at the specific moments with topological edge state. The corresponding temperature profiles are shown on the sides. (b) Schematic diagram of one domain wall consisting of five nontrivial (20 units) and five trivial (20 units) lattices. (c) The experimental locations of the maximum temperature point for each unit at the specific moments with one domain wall. The corresponding temperature profiles are shown aside. Adapted from Ref.~\cite{ZFLiu-XuNP22}. With permission from the Author.}
\label{Fig11}
\end{figure}  

Coming to the temperature profile, the locations of maximum temperature points at the specific moments are presented in Fig.~\ref{Fig11}(a) for this arrangement. In this case, the thermal profile is robust and stationary, with some slight deviations with respect to the initial position at the two boundaries of the system, which is a signature of topological edge states. Besides, one domain wall composed of five nontrivial (20 units) and five trivial (20 units) lattices can be configured to implement an interface state [see Fig.~\ref{Fig11}(b)]. The temperature field shows a robust and stationary profile in the nontrivial lattice while it shows large deviations to the initial state and becomes homogenized finally in the trivial lattice [see Fig.~\ref{Fig11}(c)]. This exotic thermal profile shows great potential for manipulating the thermal field.

Higher-order topological insulator, which is beyond the conventional bulk-boundary correspondence, arouses much attention from the community of condensed matter physics~\cite{ZFLiu-BenalcazarSci17, ZFLiu-BenalcazarPRB17, ZFLiu-SchindlerSA18, ZFLiu-BenalcazarPRB19}. Generally speaking, a $d$-dimensional $n$th-order topological insulator exhibits ($d-n$)-dimensional gapless boundary state ($n>1$). Except in condensed matter physics, the higher-order topological insulator has also been realized in the classical wave systems~\cite{ZFLiu-XieNRP21, ZFLiu-ImhofNP18, ZFLiu-HeNC20, ZFLiu-QiPRL20}. There are two classes of higher-order topological insulators in the physics community, those without multipole moments, denoted by the 2D Su-Schrieffer-Heeger model, and those with multipole moments, denoted by the Benalcazar-Bernevig-Hughes model. Though there are some works which have realized the higher-order topological insulator without multipole moment in thermal diffusion~\cite{ZFLiu-LiuarXiv22-1, ZFLiu-QiarXiv23, ZFLiu-WuAM23}, the quadrupole topological phase has not been implemented due to the absent bulk quadrupole moment and undefined negative couplings in heat transfer. A recent work has addressed these problems and realized the non-Hermitian quadrupole topological insulator in diffusion systems~\cite{ZFLiu-XuNC23}. 

\begin{figure}[!ht]
\includegraphics[width=0.9\linewidth]{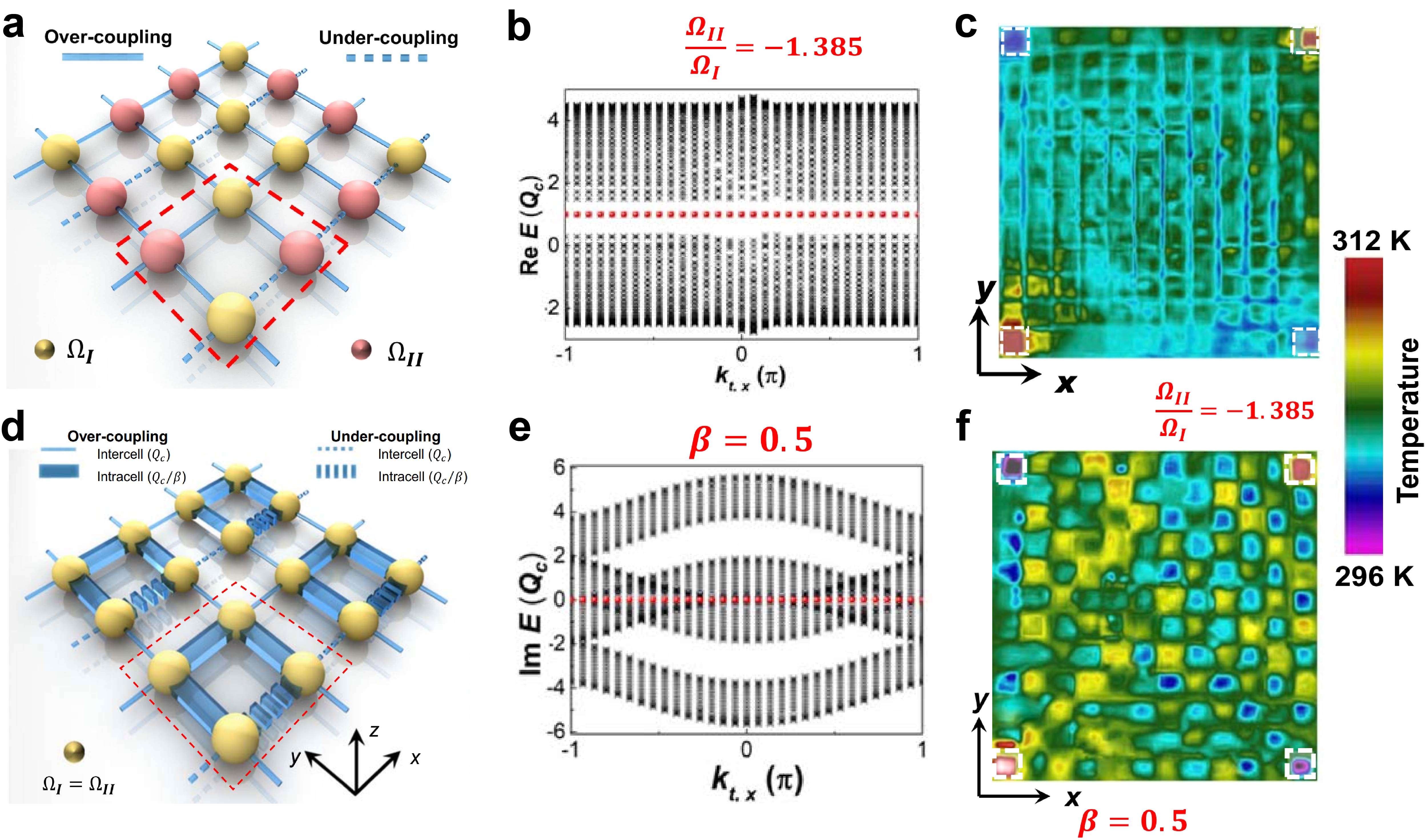}
\caption{2D quadrupole topological insulator in convective thermal metamaterials. (a) Schematic diagram of the square thermal lattice consisting of 16 sites with different advections. The red dashed box indicates a four-site unit-cell structure. $\Omega_{{\uppercase\expandafter{\romannumeral1}}/{\uppercase\expandafter{\romannumeral2}}}$ represents the magnitude of the angular velocities imposed on each site. (b) The theoretical real spectrum induced by Hermitian advection at the corner state. (c) Experimental temperature distributions at the corner state induced by Hermitian advection. (d) Schematic diagram of the square thermal lattice with different coupling strengths. The intercell and intracell channels are fabricated in different structures, enabling different thermal coupling strengths. (e) The theoretical imaginary spectrum induced by non-Hermitian couplings at the corner state. (f) Experimental temperature distributions at the corner state induced by non-Hermitian couplings. Adapted from Ref.~\cite{ZFLiu-XuNC23}. With permission from the Author.}
\label{Fig12}
\end{figure}

A convective thermal lattice with multiple discrete sites is shown in Fig.~\ref{Fig12}(a). Each site represents a finite volume heat transfer process and the grid lines between neighboring sites correspond to their thermal couplings. Then the tunable advections $\Omega_{{\uppercase\expandafter{\romannumeral1}}/{\uppercase\expandafter{\romannumeral2}}}$ are adopted on each site to form a unit-structure consisting of four neighboring sites, as shown in the red dashed box of Fig.~\ref{Fig12}(a). The neighboring sites are coupled via the heat exchanges induced by intracells and intercells. $\beta$ is denoted as the ratio between the intercell and intracell thermal coupling strengths. Two types of tilted connections (over-coupling and under-coupling) are designed to mimic the coexisting positive and negative nearest-neighbor couplings with opposite isotherms. 

With this structure, Hermitian advection and non-Hermitian couplings provide two modulations that enable a quadrupole topological phase. For the Hermitian advections, the $\Omega_{\uppercase\expandafter{\romannumeral2}}$ is adjusted while the $\Omega_{\uppercase\expandafter{\romannumeral1}}$ is kept unchanged. Besides, $\beta$ is set to be one to ensure the heat exchange areas of the intracell and intercell components are the same. In this case, the imaginary bands are gapless, so that in-gap corner states can be observed in the real spectrum. When modulating the $\Omega_{\uppercase\expandafter{\romannumeral2}}$ at $\Omega_{\uppercase\expandafter{\romannumeral2}}=-1.385\Omega_{\uppercase\expandafter{\romannumeral1}}$, the robust in-gap corner states can be found in the theoretical real band [see in Fig.~\ref{Fig12}(b)]. In order to experimentally observe these corner states, a thermal system consisting of ${\rm{12{\times}12}}$ sites is fabricated. All of the sites are hollowed out to allow water to pass. The measured temperature profile at the corner state is shown in Fig.~\ref{Fig12}(c), with a relatively high field intensities at four corners. Then the non-Hermitian thermal couplings are modulated as shown in Fig.~\ref{Fig12}(d). With the same advections on all sites and intercell coupling channels, the intracell coupling channels are enlarged by inserting internal fins to modulate the $\beta$. In this case, the real-valued band is gapless and the imaginary-valued band is gapped. At $\beta=0.5$, the 0D corner states exist in the imaginary spectrum [see in Fig.~\ref{Fig12}(e)]. The experimental temperature field distribution for the corner state is shown in Fig.~\ref{Fig12}(f). The higher-order topological state can be observed both in the real-valued (modulated by advections) and imaginary-valued (modulated by thermal couplings) bands, which breaks the restriction that only real-valued band can exhibit corner states in classical wave systems.

Besides, a separate work exhibits the implementation of a Chern insulator with the chiral edge state in diffusion systems~\cite{ZFLiu-XuEPL21-2}. The basic unit is a hexagonal structure with thermal convection. Based on this structure, a honeycomb lattice can be designed to mimic the Haldane model~\cite{ZFLiu-HaldanePRL88}. A steady temperature field will propagate unidirectionally along the boundary of the system without backscattering.

\section{${\rm{\textbf{\uppercase\expandafter{\romannumeral5}}}}$. Conclusion and outlook}

As a conclusion, we discuss some interesting works on non-Hermitian physics and topological phenomena in convection-conduction systems. From the point of view of non-Hermitian physics, we first discuss the realization of EP in thermal diffusion, followed by the high-order EP and the dynamically encirclement of EP. As two extensions of the EP, we introduce chiral heat transport around the EP and diffusive Weyl exceptional ring. From topological physics, 1D thermal topological insulators and 2D thermal quadrupole topological insulators have been introduced as two examples of diffusive topological phases. 

The promising future of topological and non-Hermitian convective thermal metamaterials can be anticipated. Until now, only a small fraction of exotic phases of matter have been realized in diffusive systems. The realization of a wide variety of topological states is a prevailing trend in this field. From the non-equilibrium dynamics, except for non-Hermitian physics, Floquet thermal metamaterials can be devised by spatiotemporal modulating the materials’ parameters~\cite{ZFLiu-Yinelight22}. From the higher-dimensional systems, many peculiar configurations, such as fractal systems~\cite{ZFLiu-BiesenthalSci22}, moir${\rm{\acute{e}}}$ lattices~\cite{ZFLiu-WangNat20}, and disclinations~\cite{ZFLiu-LiuNat21}, can be engineered in the thermal lattices. The nonlinearity can also be introduced into non-Hermitian and topological thermal metamaterials to explore more interesting physics~\cite{ZFLiu-FanJPCC2009, ZFLiu-HuangJRCC08, ZFLiu-ZhuJAP08, ZFLiu-HuangNY07, ZFLiu-WangAPL07, ZFLiu-TianCPL06, ZFLiu-FanAPL06, ZFLiu-XuPLA06, ZFLiu-HuangPR06a, ZFLiu-WangCPL06, ZFLiu-TianPRE06, ZFLiu-HuangJAP06, ZFLiu-HuangAPL05a, ZFLiu-HuangJOSAB05, ZFLiu-HuangJPCB05, ZFLiu-HuangOL05, ZFLiu-HuangAPL05b, ZFLiu-HuangPLA05, ZFLiu-HuangPRE04c, ZFLiu-HuangJCP04, ZFLiu-HuangJPCM04, ZFLiu-HuangEL04, ZFLiu-HuangPRE04e, ZFLiu-HuangAPL04, ZFLiu-HuangPRE04g, ZFLiu-GaoPRB04, ZFLiu-GaoEPJB03, ZFLiu-HuangJAP03, ZFLiu-HuangPRE01, ZFLiu-PanPB01, ZFLiu-HuangCTP01-1, ZFLiu-HuangSSC00, ZFLiu-HuangJAP05}. In addition, machine learning and other algorithms are expected to be valuable in studying topological effects in diffusion systems~\cite{ZFLiu-ZhangPRL18, ZFLiu-ZhangCPL23, ZFLiu-ZhangPRD22, ZFLiu-LiuJAP21-1, ZFLiu-JinIJHMT21, ZFLiu-LiuJAP21-2}. The econophysics may inspire the study of non-Hermitian and topological convective thermal metamaterials~\cite{ZFLiu-XinFP17, ZFLiu-XinPA17, ZFLiu-JiPA18, ZFLiu-Tan2, ZFLiu-Tan5, ZFLiu-Tan6, ZFLiu-Tan9, ZFLiu-Tan12, ZFLiu-Tan13, ZFLiu-QiuPLA15, ZFLiu-QiuPR15, ZFLiu-QiuSpringer15, ZFLiu-QiuPLoS14-1, ZFLiu-QiuJSM14, ZFLiu-QiuCPB14, ZFLiu-QiuPLoS14-2, ZFLiu-QiuPLA14, ZFLiu-QiuFP14, ZFLiu-QiuPLoS13, ZFLiu-QiuEPJB13, ZFLiu-Liang2013fopw, ZFLiu-wei2013AnAM, ZFLiu-Liu2013ASP, ZFLiu-liang2013pre, ZFLiu-wei2013plos, ZFLiu-li2012epl, ZFLiu-SongARCS12, ZFLiu-ZhaoPNAS11, ZFLiu-WangPNAS09, ZFLiu-GuCTP2009, ZFLiu-ZhouPA2009, ZFLiu-YePA081, ZFLiu-ChenJPA07}. Besides, this field can also be generalized to nanoparticles~\cite{ZFLiu-ZhouEPL23, ZFLiu-LinSCPMA22, ZFLiu-HuangAMT22, ZFLiu-GaoNM21, ZFLiu-MengCPB18, ZFLiu-Tan3, ZFLiu-Tan7, ZFLiu-Tan8, ZFLiu-Tan10, ZFLiu-Tan14, ZFLiu-Tan17, ZFLiu-Tan18, ZFLiu-Tan19, ZFLiu-QiuJP.Phys.Chem.B15, ZFLiu-QiuTEPJ-AP14, ZFLiu-Meng2013pre, ZFLiu-Fan2013cpb, ZFLiu-Wang2013ACM, ZFLiu-Meng2013mp, ZFLiu-Wang2012cpb, ZFLiu-WangJPCB11, ZFLiu-MengJPCB11, ZFLiu-BaoJPCM10, ZFLiu-TanSM10, ZFLiu-XiaoPRB05}, complex fluids~\cite{ZFLiu-QiuCTP15, ZFLiu-chen2013prl, ZFLiu-Li2013EPJP, ZFLiu-Fan2013fop, ZFLiu-Li2012sm, ZFLiu-FanJPDAP11, ZFLiu-LiuIJMPB11, ZFLiu-GaoPP10, ZFLiu-GaoPRL10, ZFLiu-WuEPJAP09, ZFLiu-TanJPCB2009-1, ZFLiu-XiaoJPCB2008, ZFLiu-FanJAP2008, ZFLiu-JianJRCB08, ZFLiu-GaoJPCC07, ZFLiu-CaoJPCB06, ZFLiu-ShenCPL06, ZFLiu-HuangPRE05b, ZFLiu-HuangPRE04a, ZFLiu-HuangCP04, ZFLiu-HuangJMMM05}, and other electric behaviours~\cite{ZFLiu-FanCTP10, ZFLiu-JianJPCC2009, ZFLiu-TianJAP2009, ZFLiu-TanJPCB2009-2, ZFLiu-ZhaoJAP2009, ZFLiu-ZhangCPL2009, ZFLiu-WangOL2008, ZFLiu-WangOL2008, ZFLiu-2008, ZFLiu-GaoAPL2008, ZFLiu-ZhangAPL08, ZFLiu-ZhangAPL08, ZFLiu-XuJMR08, ZFLiu-FangCPL07, ZFLiu-TianPRE07, ZFLiu-FanJPCB06, ZFLiu-HuangPRE05a, ZFLiu-LiuCTP05, ZFLiu-HuangPRE04b, ZFLiu-HuangPLA04, ZFLiu-HuangJPCB04, ZFLiu-KoEPJE04, ZFLiu-HuangPRE04f, ZFLiu-HuangCPL04, ZFLiu-DongJAP04-1, ZFLiu-LiuPLA04, ZFLiu-KoJPCM04, ZFLiu-DongJAP04-2, ZFLiu-HuangPRE03-1, ZFLiu-GaoPRE03, ZFLiu-HuangPRE03-2, ZFLiu-HuangCTP03, ZFLiu-HuangPLA02, ZFLiu-HuangCTP02, ZFLiu-HuangPRE02, ZFLiu-HuangJPCM02, ZFLiu-HuangCTP01-2}. Potential applications in topological convective thermal materials are also to be expected. Topological edge states may be useful for localized heat management and robust heat transport. In brief, topological and non-Hermitian convective thermal metamaterials are highly interdisciplinary and of great scientific and technological values.

\end{document}